\documentclass[pra,preprint,showpacs]{revtex4}
\usepackage{graphicx}
\usepackage{amsmath}
\newcommand{\be}{\begin{equation}}
\newcommand{\ee}{\end{equation}}
\begin{document}

\title{Bulk and surface localized modes on a magnetic nonlinear impurity}

\author{Mario I. Molina}

\affiliation{Departamento de F\'{\i}sica, Facultad de Ciencias,
Universidad de Chile, casilla 653, Santiago, Chile}


\begin{abstract}
We study localized modes on a single magnetic impurity positioned in the bulk or at the surface of a one-dimensional 
chain, in the presence of a magnetic field $B$ acting at the impurity site.  The strong on-site nonlinear interaction 
$U$ between two electrons of opposite spin at the impurity site, modelled here as a nonlinear local term, 
and the presence of the external field induce a strong correlation between parallel and anti-parallel spin bound  states. 
We find that, for an impurity in the bulk, a localized vector mode (with up and down spin components) is always possible 
for any given value of  $U$ and $B$,  while for a surface impurity, a minimum value of both, $U$ and $B$ is needed to 
create a vector mode.  In this case, up to two localized  modes are possible, but only one of them is stable. 
The presence of the surface  seems to destabilize the bulk mode  in the parameter region $U \sim B$, creating a 
``forbidden strip'' region in  parameter space, bounded by   $U=B+V$ and $U=B-V$, approximately.

\end{abstract}


\maketitle

\section{Introduction}
Nonlinear effects are the subject of intensive, ongoing research in several fields that include  
molecular crystals\cite{Sievers-Takeno, Kivshar-Campbell}, Josephson-junction arrays\cite{junctions}, 
Ferromagnetic materials\cite{ferro}, photonic crystals\cite{pc}, photonic lattices\cite{pl}, 
Bose-Einstein condensates in magneto-optical traps\cite{bec} and nonlinear metamaterials\cite{meta}, to name
a few.
This interest is due in no small part to the wide range of potential applications to the design and operation 
of optoelectronic devices.

In condensed matter, nonlinear effects arise from at least two different sources. One
of them is a possible strong coupling between an excitation and local vibrational modes. In the approximation
where one assumes a rapid readjustments of the local vibrations to the presence of the excitation, one
quickly arrives at some version of the discrete nonlinear Schr\"{o}dinger  (DNLS) equation, whose exact
form depends on the type of anharmonicity (or lack of it) of the underlying oscillators\cite{adiabatic}. When the
strong interaction tales place only at some few sites, or at a single impurity site, it is possible to compute the
formation of bound states in closed form in one-dimensional\cite{uno}, two-dimensional\cite{dos} and 
three-dimensional lattices\cite{tres}, by using a direct extension of the lattice Green function formalism. 

The other source for nonlinearity comes from electron-electron interactions in nanoscale devices, such as quantum dots
and few impurity models\cite{Johnson_jpcm}. Roughly speaking, the Coulomb interaction gives rise to
a nonlinear term in the Schr\"{o}dinger equation, that is usually modelled as a cubic, nonlocal term in
the fermionic field operators. This is usually followed by the Hartree-Fock approximation for the nonlinear 
term\cite{hartree}, or the use of a perturbative approach in the Coulomb interaction\cite{perturbative}.
Another recent approach calls for modelling the electron-electron interaction by a nonlinear local term in 
the Schr\"{o}dinger equation\cite{bahlouli_molina}. While being a sort of oversimplification of the many-body
problem, it has the advantage of retaining the main features of Coulomb interaction-induced nonlinearities, 
while allowing for speedy computation of quantities of interest for electronic transport. This approach has been
recently used for a simple computation of the zero-voltage conductance across a magnetic 
impurity\cite{bahlouli_molina}.

In this Letter, we focus on the possible localized impurity modes that can reside on top of a nonlinear 
magnetic impurity, where the source of the nonlinearity is due to strong electron-electron interaction effects at
the impurity site. It is important to ascertain the conditions under which such bound states exist, since they could,
for instance, scatter other extended excitations in the system. The precise control of this scattering could be of 
importance in the control of the transport properties across nanoscale impurity regions.

\section{Model}
Let us start by considering the problem of a one-dimensional discrete system consisting
of two linear chains (leads) joined by a strongly nonlinear magnetic region, where electron-electron
effects are important. In the interacting region, the electron-electron repulsion is modelled by
$U \rho_{0 \sigma} \rho_{0 -\sigma}$, where $\rho_{0 \sigma}=|\psi_{0,\sigma}|^2$, where $\psi_{0,\sigma}$ 
is the probability amplitude of finding an electron of spin $\sigma=\uparrow,\downarrow$ on site $n=0$.

The coupled evolution equations for the probability amplitudes are
\begin{eqnarray}
i {d \psi_{n,\uparrow}\over{d t}} + V (\psi_{n+1,\uparrow} + \psi_{n-1,\uparrow}) + 
\delta_{n,0}  (\epsilon_{0,\uparrow}+ U |\psi_{n,\downarrow}|^{2}) \psi_{n,\uparrow} & = & 0\nonumber\\
i {d \psi_{n,\downarrow}\over{d t}} + V (\psi_{n+1,\downarrow} + \psi_{n-1,\downarrow}) + 
\delta_{n,0}  (\epsilon_{0,\downarrow}+ U |\psi_{n,\uparrow}|^{2}) \psi_{n,\downarrow} & = & 0\label{eq:1}
\end{eqnarray}
where $V$ is the nearest-neighbor coupling 
parameter, $U$ is the Coulomb repulsion energy and $\epsilon_{0,\uparrow} = -B,\epsilon_{0,\downarrow} = B $ is the 
Zeeman energy shift due to the magnetic field $B$, which is assumed to be appreciable only in the 
immediate vicinity of the impurity site.

We are interested in stationary solutions of the type $\psi_{n,\sigma}(t) = \exp(i \beta t)\ \psi_{n,\sigma}$.
This leads to the system of nonlinear equations
\be
-\beta \psi_{n,\sigma} + V (\psi_{n+1,\sigma} + \psi_{n-1,\sigma}) + 
\delta_{n,0}  (\epsilon_{0,\sigma}+ U |\psi_{n,-\sigma}|^{2}) \psi_{n,\sigma} = 0.\label{eq:2}
\ee
At this point an general observation is in order. If we were to consider the impurity site also coupled to a local, 
fast elastic vibrational degree of freedom, then in the
limit when the local vibration is completely enslaved to the electronic motion, there would be an additional term of the
form $\gamma |\psi_{n,\sigma}|^{2} \psi_{n,\sigma}$ in  Eqs.(\ref{eq:1}), (\ref{eq:2}). In that case, the system would 
be formally equivalent to a birefringent, nonlinear optical Kerr impurity embedded in a weakly-coupled,  linear waveguide 
array,  in the absence of four-wave  effects, and $\psi_{n,\uparrow}$ and  $\psi_{n,\downarrow}$ would represent the 
electric field amplitudes for the  TM and TE mode, respectively\cite{stegeman}. 

In order to keep the magnetic and electronic repulsion effects well separated from possible polaronic effects, 
in this work we restrict ourselves to Eqs. (\ref{eq:1}), (\ref{eq:2}). The interplay of external magnetic field, 
electron-electron repulsion and electron-phonon interaction effects on a single impurity, will be described elsewhere.  

\section{Impurity in Bulk}

Here the chain occupies the interval $-\infty<n<\infty$, with the impurity site at $n=0$. 
We look for a localized mode centered on the impurity site, $\psi_{n,\sigma}= A_{\sigma}\ \xi_{\sigma}^{|n|}$,
with $0<|\xi|<1$. This {\em ansatz} leads to the equations:
\be
-\beta + 2 V \xi_{\sigma} + \epsilon_{\sigma} + U A_{-\sigma}^{2} = 0\label{eq:3}
\ee
and
\be
\beta = V \left( \xi_{\sigma} + {1\over{\xi_{\sigma}}} \right). \label{eq:4}
\ee 

On the other hand, from probability conservation, 
we have $1 = \sum_{n} |\psi_{n,{\uparrow}}|^{2} = \sum_{n} |\psi_{n,{\downarrow}}|^{2} $. This implies the additional
equations
\be
1 = A_{\uparrow}^{2} \left( {1 + \xi_{\uparrow}^{2}\over{1 - \xi_{\uparrow}^{2}}} \right)\label{eq:5}
\ee
and
\be
1 = A_{\downarrow}^{2} \left( {1 + \xi_{\downarrow}^{2}\over{1 - \xi_{\downarrow}^{2}}} \right).\label{eq:6}
\ee
\begin{figure}[t]
\noindent\includegraphics[scale=0.75]{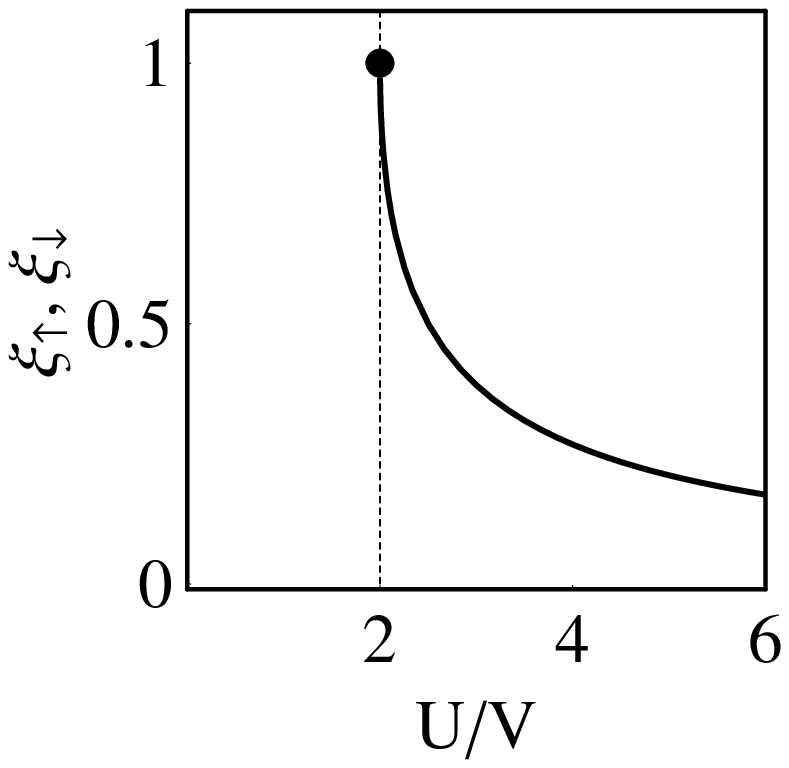}\hspace{0cm} 
\noindent\includegraphics[scale=0.75]{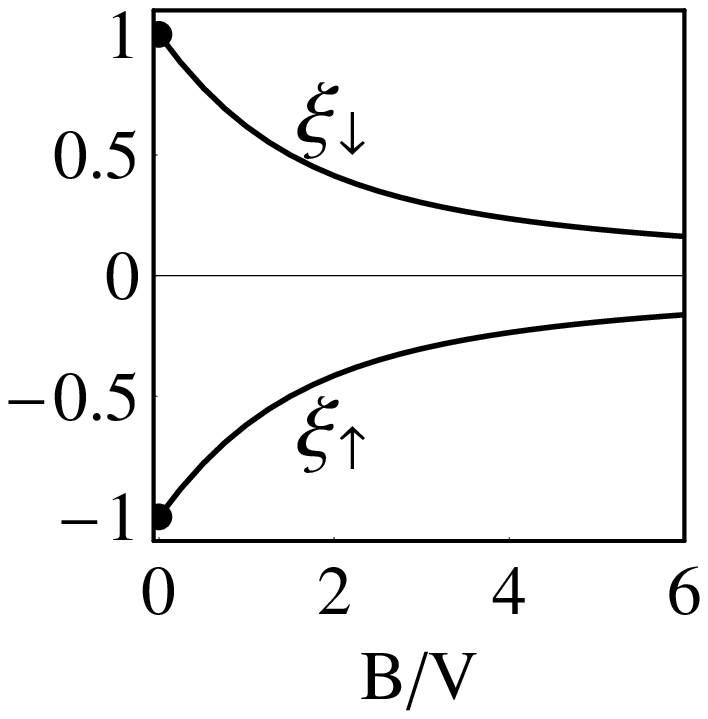}
\caption{(Color online). Left: $B=0$ case. Amplitude of localized state for both spins as a function of nonlinearity.
Right: $U=0$ case. Amplitude of localized state for both spins, as a function of magnetic field.  }
\label{newfig1}
\end{figure}
From Eqs.(\ref{eq:3}),(\ref{eq:4}),(\ref{eq:5}) and (\ref{eq:6}), we obtain two coupled equations for $\xi_{\sigma}$:
\be
\xi_{\uparrow}^{2} = 1 - \xi_{\uparrow} \left[ -{B\over{V}} + {U\over{V}} \left( {1-\xi_{\downarrow}^{2}\over{1+\xi_{\downarrow}^{2}}} \right) \right]\label{eq:7}
\ee
and
\be
\xi_{\downarrow}^{2} = 1 - \xi_{\downarrow} \left[ +{B\over{V}} + {U\over{V}} \left( {1-\xi_{\uparrow}^{2}\over{1+\xi_{\uparrow}^{2}}} \right) \right].\label{eq:8}
\ee
It is clear from these equations that, the transformation $\xi_{\uparrow}\rightarrow \xi_{\downarrow}$, $B\rightarrow -B$ leave the equations
invariant. Since $U>0$ due to the repulsive nature of the Coulomb interaction, we only need to deal with, say, $B>0$. 
The behavior for the opposite sign of $B$ is obtained by simply exchanging $\xi_{\uparrow}$ and $\xi_{\downarrow}$. 

Before analyzing the general case, let us discuss briefly a couple of important special cases.

(a) {\bf B=0}. In this case, there is no physical distinction between $\xi_{\uparrow}$ and  $\xi_{\downarrow}$,
and Eqs.(\ref{eq:7}) and (\ref{eq:8}) collapse into a single one: $\xi^{2} =1 + (U/V) \xi ((1-\xi^{2})/(1+\xi^{2}))$, 
with solution
\be
\xi = \left({U\over{2 V}} \right) - \sqrt{\left({U\over{2 V}}\right)^{2}-1},
\ee
provided $U/V > 2$. Since $\xi<0$, the spatial mode profile is staggered. Figure 1 shows $\xi$ in terms of
$U/V$. As expected, an increase of nonlinearity reduces the width of the localized mode. 
\begin{figure}[t]
\noindent
\includegraphics[scale=0.65]{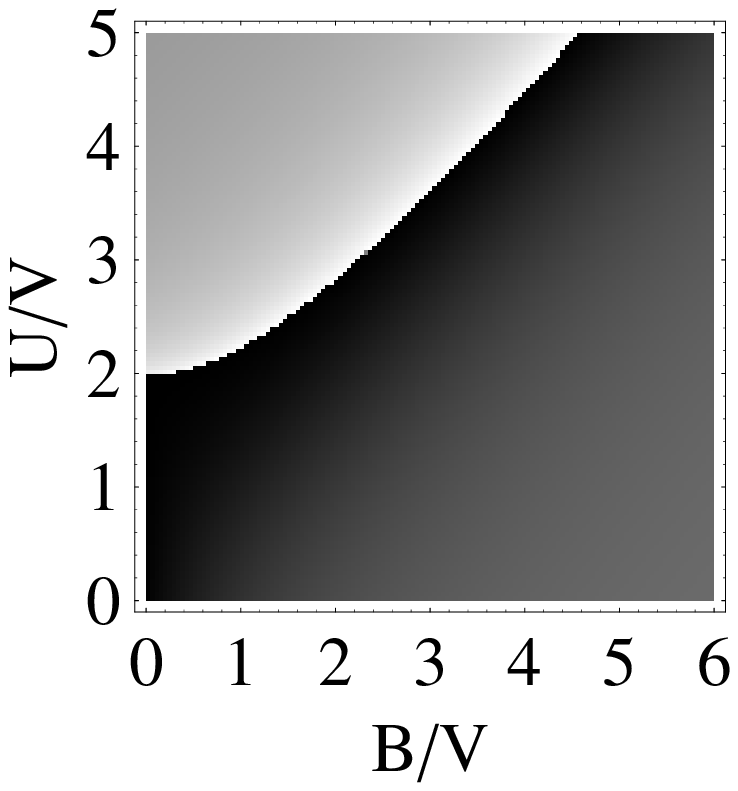}
\includegraphics[scale=0.65]{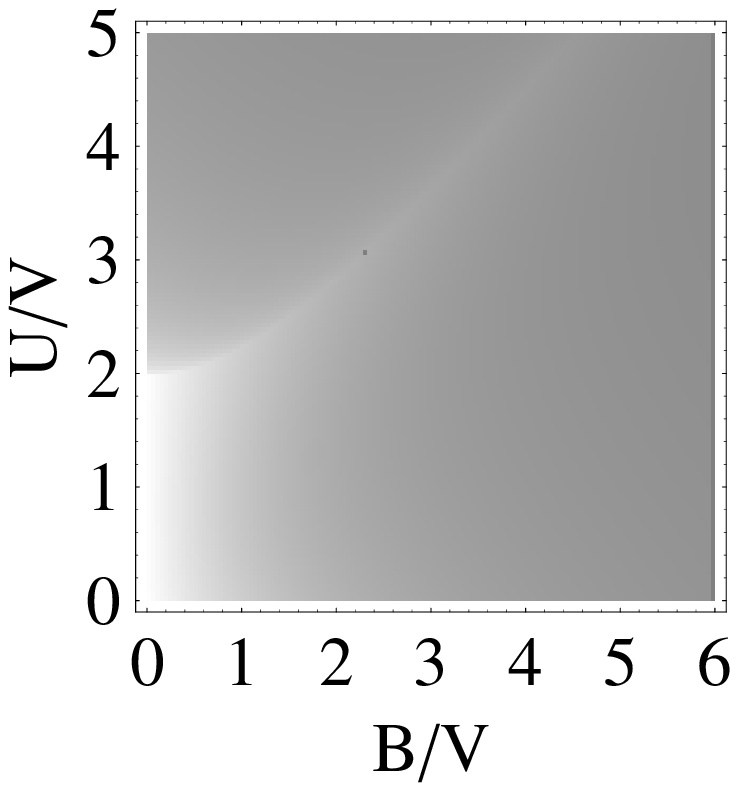}
\caption{Amplitude of localized mode as a function of nonlinearity and
external magnetic field, for spin up (left) and spin down (right). Darkest (whitest) shade 
corresponds to $\xi=-1(1)$. }
\label{figura2} 
\end{figure}

(b)\ {\bf U=0}. In the absence of Coulomb repulsion,  the only source for localization is given by the
presence of the external field $B$. The equations read now $\xi_{\uparrow}^{2} = 1 - (B/V) \xi_{\uparrow},
 \xi_{\downarrow}^{2} = 1 + (B/V) \xi_{\downarrow}$, with solutions
 \be
 \xi_{\uparrow} = \left({B\over{2 V}}\right) - \sqrt{\left({B\over{2 V}}\right)^{2}+1}
\ee
and
 \be
 \xi_{\downarrow} = -\left({B\over{2 V}}\right) + \sqrt{\left({B\over{2 V}}\right)^{2}+1}.
\ee
Figure 1 shows the allowed $\xi_{\uparrow},\xi_{\downarrow}$ as a function of magnetic field. In this case,
no minimum field strength is needed to create a localized mode, but the mode profile corresponding to 
a spin parallel (antiparallel) to the external field is staggered (unstaggered).

For the general case, we solve Eqs.(\ref{eq:7}) and (\ref{eq:8}) numerically, selecting the real roots that
lie in the interval $(-1,1)$. Results are shown on Fig.2 in the form of a nonlinearity-magnetic field phase
space diagram, showing the amplitude of $\xi_{\uparrow}, \xi_{\downarrow}$, for a given value of $U/V$ and $B/V$.
\begin{figure}[t]
\noindent
\includegraphics[scale=0.6]{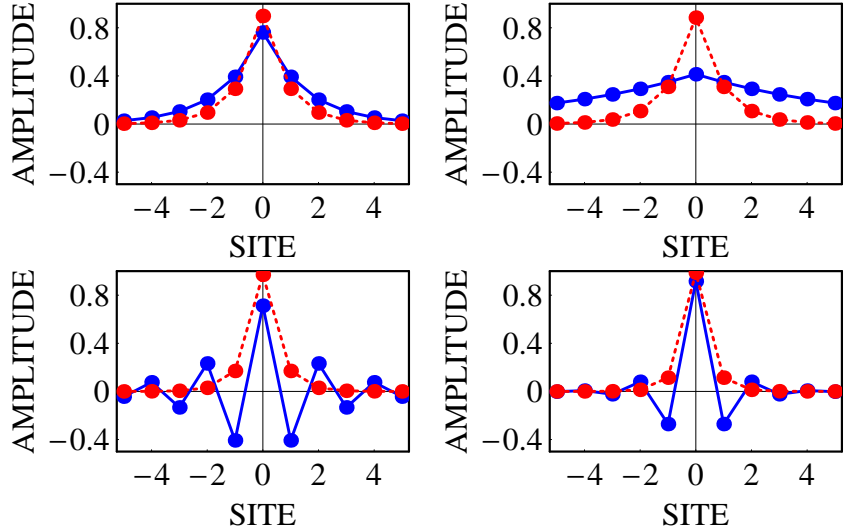}
\caption{(Color online). Spatial mode profiles for $U/V=3$ and varying values
of $B/V$. Top left: $B/V=1$. Top right: $B/V=2$. Bottom left: $B/V=4$. Bottom right: $B/V=6$. 
Solid (dashed) curves denote the spin up (down) mode. }
\label{figura3}
\end{figure}
For spin up, the amplitude is always negative and increasing with increasing $B$, provided $U/V<2$.  When $U/V>2$,
there is a sharp boundary in $U-B$ space separating positive amplitude values, in the ``low'' $B/V$ region, from negative 
amplitude values, in the ``high'' $B/V$ value. Upon crossing the boundary, in the direction of increasing $B$, 
the sign of the amplitude jumps  discontinuously from $+1$ to $-1$. The shape of this critical boundary can be 
computed exactly from Eqs.(\ref{eq:7}) and (\ref{eq:8}): We set $\xi_{\uparrow}^{2}=1$ in Eq.(\ref{eq:8}), and solve for
$\xi_{\downarrow}=-(B/2V)+\sqrt{(B/2V)^2+1}$. Next, we insert this into Eq.(\ref{eq:7}), and solve for $U$ in terms of $B$,
obtaining:
\be
\left({B\over{V}}\right)_{c} = {4 + (B/V)^{2}-(B/V)\sqrt{(B/V)^2+4}\over{-(B/V)+\sqrt{(B/V)^2+4}}}. \label{eq:12}
\ee

For spin down, the amplitude of the localized modes is always positive. For $U/V<2$, the amplitude 
decreases monotonically towards zero  with an increase in $B$. For $U/V>2$ however, the amplitude starts
at some positive value at $B=0$, then increases further with an increase in B, until $B/V$ reaches the 
critical boundary computed above. After that, an increase in $B$ leads to an decrement of the amplitude towards
zero.

In Fig.3 we show some spatial profiles $\psi_{n,\sigma}$ for a fixed value of nonlinearity, 
and increasing values of magnetic field. Note that for $B>0$, the localized mode for spin down is always unstaggered,
while for spin up, the mode is unstaggered at first, and becomes staggered after certain value of magnetic field, given 
by Eq.(\ref{eq:12}).

\section{Surface impurity}
We now consider the case when the magnetic impurity is at the very beginning of a semi-infinite lattice. 
We relabel the previous chain, so that the first site is now at $n_{0}=0$. The 
stationary-state equations read now
\be 
-\beta\ \psi_{0,\sigma} + V \psi_{1,\sigma} + (\epsilon_{0,\sigma} + U |\psi_{0,\sigma}|^{2}) \psi_{0,\sigma} = 0,\label{eq:13} \ee
\be
-\beta\ \psi_{n,\sigma} + V ( \psi_{n+1,\sigma} + \psi_{n-1,\sigma} ) = 0,\hspace{1cm}n=0,1,2,\ldots
\label{eq:14}
\ee

We proceed as before and pose a solution of the form $\phi_{n,\sigma} = A_{\sigma}\xi_{\sigma}^{n}$, 
where $0<|\xi|<1$ and $n=0,1,2,\dots$. After replacing
this ansatz into Eq.(\ref{eq:13}) and (\ref{eq:14}), one obtains
$\beta=V \xi_{\sigma}+\epsilon_{0,\sigma}+U A_{-\sigma}^{2}$ and $\beta=\xi_{\sigma}+(1/\xi_{\sigma})$, 
which implies
\be%
\xi_{\sigma} = {1\over{(\epsilon_{\sigma}/V)+ (U/V) A_{-\sigma}^{2}}}.\label{eq:15}
\ee
On the other hand, from the normalization condition, 
$1=\sum_{n} |\psi_{n,{\uparrow}}|^{2}=\sum_{n} |\psi_{n,{\downarrow}}|^{2}$, we obtain
$A_{\uparrow}^{2}=1-\xi_{\uparrow}^{2}$ and $A_{\downarrow}^{2}=1-\xi_{\downarrow}^{2}$.
After inserting this into Eq.(\ref{eq:15}) and after using $\epsilon_{0,\sigma}=-\sigma B$,
one arrives at two coupled transcendental equations for $\xi_{\uparrow}, \xi_{\downarrow}$:
\be
\xi_{\uparrow} = {1\over{-(B/V)+(U/V)\left(1-\xi_{\downarrow}^{2}\right)}},
\ee
\be
\xi_{\downarrow} = {1\over{(B/V)+(U/V)\left(1-\xi_{\uparrow}^{2}\right)}}.
\ee
As was done for the impurity in the bulk, let us consider first two special cases.

(a) {\bf B=0}. In this case, $\xi_{\uparrow}=\xi_{\downarrow}=\xi$, where $\xi$ satisfies
$\xi(1-\xi^{2})=(U/V)^{-1}$. This implies that a critical value $(U/V)_{c}=(3/2)\sqrt{3}\sim 2.6$ exists, such that for
$(U/V)<(U/V)_{c}$, no localized state exists. For $(U/V)=(U/V)_{c}$, there is exactly one localized
state, and $(U/V)>(U/V)_{c}$, two localized states are possible; for one of them, $\xi$ increases towards
unity as nonlinearity is increased (unstable mode), while the other decreases $\xi$ as nonlinearity is 
increased (stable mode). Since $U>0$, this implies that $\xi$ is also positive, and thus, the localized mode
is unstaggered. Figure 4 shows $\xi$ in terms of
$U/V$.

(b)\ {\bf U=0}. In the absence of Coulomb repulsion, we immediately obtain $\xi_{\uparrow}=-1/(B/V)$ and
$\xi_{\downarrow}=1/(B/V)$, where $|B/V|>1$ to ensure $|\xi_{\uparrow,\downarrow}|<1$.
Figure 4 shows the allowed $\xi_{\uparrow},\xi_{\downarrow}$ as a function of magnetic field. 
The mode profile corresponding to  a spin parallel (antiparallel) to the external field is staggered (unstaggered).
\begin{figure}[t]
\noindent\includegraphics[scale=0.75]{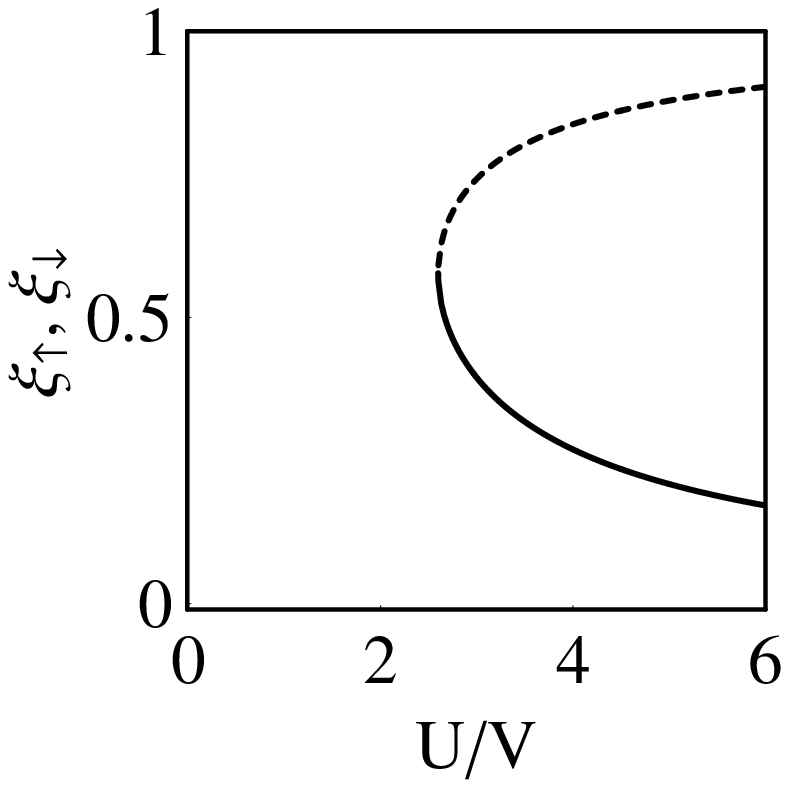}\hspace{0cm} 
\noindent\includegraphics[scale=0.75]{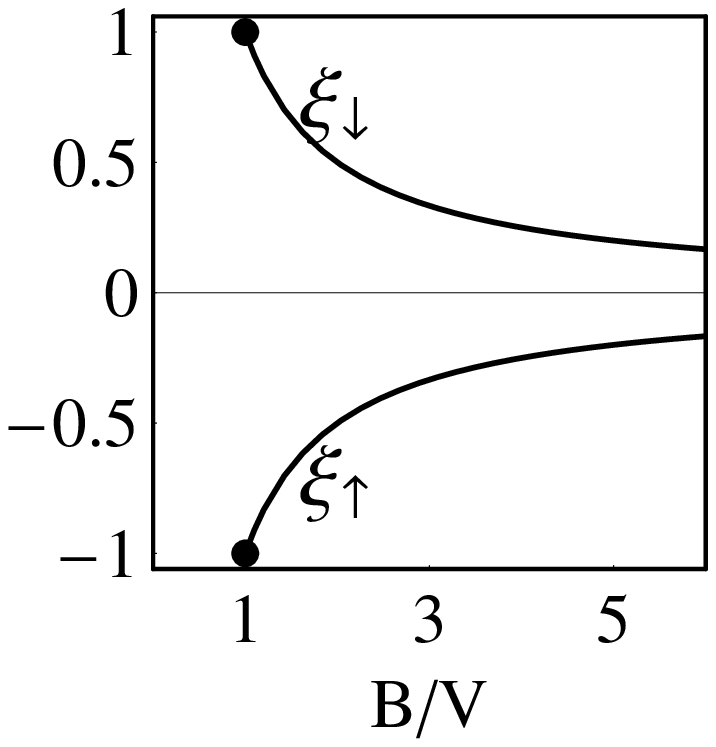}
\caption{Left: $B=0$ case. Amplitude of localized state for both spins as a function of nonlinearity.
The solid (dashed) curve denotes the stable (unstable) mode. 
Right: $U=0$ case. Amplitude of localized state for both spins, as a function of magnetic field.  }
\label{fig4}
\end{figure}

Although Figs. 1 and 4 look qualitatively similar, we note some interesting differences:
For the surface impurity, and in the absence of magnetic field, the amount of nonlinearity 
needed to effect a localized state is {\em higher} that in the bulk impurity case. Also, in the linear case ($U=0$),
one needs {\em higher} values of magnetic field to create a surface localized mode. The surface, or boundary
of the system is acting in a {\em repulsive} manner. 

For the general case, there are two important regimes, $B/V \leq 1$ and $B/V >1$.

{\bf $B/V \leq 1$}:\ In this case, there is a critical nonlinearity value $(U/V)_{c}$ such that, for
$0<(U/V)<(U/V)_{c}$, no localized state exists, while for $(U/V)>(U/V)_{c}$, there are two localized 
states. One of them (stable mode) becomes narrower with increasing nonlinearity; the other (unstable mode) 
becomes wider.

{\bf $B/V > 1$}:\ In this case, we have three critical values for $(U/V)$, that create four
regions:

\noindent
(i)\ $0<(U/V)<(U/V)_{c1}$:\ This is the region of ``small'' nonlinearity, where magnetic field dominates and
there is always a localized mode, with $\xi_{\uparrow}<0, \xi_{\downarrow}>0$. As $(U/V)\rightarrow (U/V)_{c1}$, 
$\xi_{\uparrow}\rightarrow -1$.\\
(ii) $(U/V)_{c1}<(U/V)<(U/V)_{c2}$:\ No localized mode exists.\\
(iii) $(U/V)_{c2}<(U/V)<(U/V)_{c3}$:\ Two localized modes are possible here. One of them (stable) becomes narrower with
increasing $U/V$, while the other (unstable) becomes wider with increasing $U/V$.\\
(iv) $(U/V)>(U/V)_{c3}$:\ The previous unstable state disappears, and only stable mode remains. As 
$(U/V)\rightarrow \infty$, both $\xi_{\uparrow}$ and $\xi_{\downarrow}$ approach zero.

Figure 5 shows all these regions in the form of a nonlinearity-magnetic field 
phase space diagram. The number inside each region denotes the number of localized
surface modes (each one with a spin up and spin down component). 
We note, in particular, the existence of a region, roughly delimited by the straight lines
$(U/V)=(B/V)+1$ and $(U/V)=(B/V)-1$, where no localized states exists. These are obtained
as limiting cases of the curves that separate regions "1" (top) and "0", and regions "0" and "1" (bottom).
For the first case the curve is $U/V=[(B/V)+1]/[1-(V/B)^{2}]$, while for the second case, the curve is
given by $U/V = [(B/V)-1]/[1-(V/B)^{2}]$. In Fig.6 we show examples of spatial profiles $\psi_{n,\sigma}$.
We nota that their staggered/unstaggered character is rather similar to the ``bulk'' case (Fig.3).

\begin{figure}[t]
\noindent\includegraphics[scale=0.8]{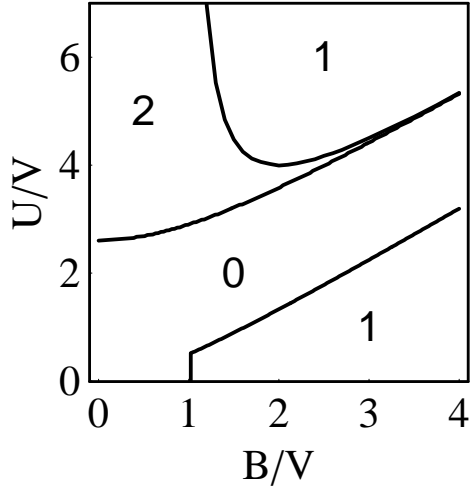}
\caption{Number of localized surface modes, as a function of nonlinearity and
magnetic field.}
\label{fig5}
\end{figure}

\begin{figure}[h]
\noindent\includegraphics[scale=0.55]{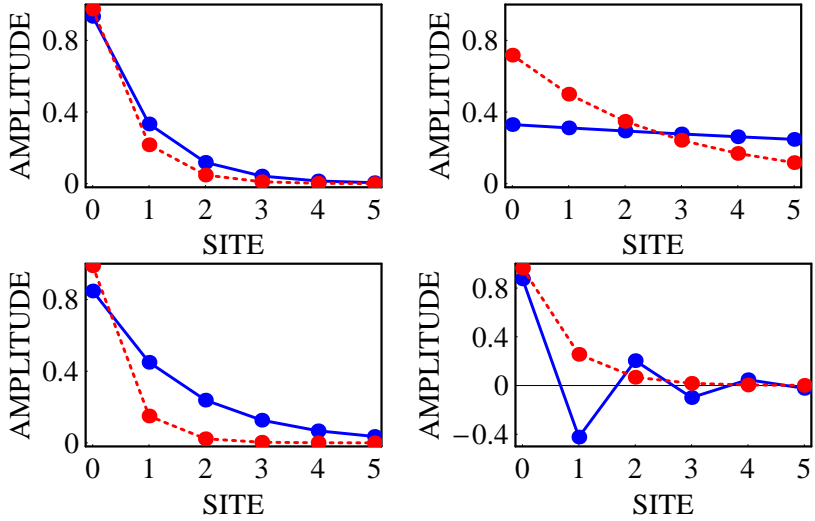}
\caption{(Color online). Examples of surface localized modes. Solid (dashed) lines refer to spin up (down). 
Top left: $B/V=1, U/V=4$ (stable mode). Top
right: $B/V=1,U/V=4$ (unstable mode). Bottom left: $B/V=3, U/V=5$. Bottom right: $B/V=3, U/V=1$.}
\label{fig6}
\end{figure}

\section{Selftrapping dynamics}

We consider now the dynamical excitation of a localized mode on top of the magnetic impurity, comparing the bulk
and surface cases. We place initially both electrons (with opposite spins) on the impurity site and follow the evolution of their probability densities,
according to Eq.(\ref{eq:1}). In particular, we focus on the time-averaged selftrapped fraction remaining the 
initial site, defined by
\be
\langle P_{\sigma} \rangle = {1\over{T}} \int_{0}^{T} |\xi_{\sigma}(t)|^{2}\ dt,
\ee
and examine how $\langle P_{\sigma}\rangle$ depends on electron interaction $U$ and magnetic field $B$.

Figures 7 and 8 show the selftrapping results for the ``bulk'' and surface cases. While the behavior in both cases
is qualitatively similar, we note that in general, it is a bit easier to selftrap in the bulk case than in the surface case.
On the other hand, for the surface case, there are sharper boundaries separating the untrapped from the selftrapped
regime. Also, in both cases and for positive $B$, selftrapping for the spin up component, is largely inhibited 
in a wide strip around the region $U\sim B$, while the spin down component is dominated by magnetic field effects.
Of course, for $U<<B$, nonlinearity effects are not important and selftrapping increases gradually with magnetic field, 
as one expects for a linear site impurity. On the contrary, for $U>>B$, selftrapping is abrupt and its 
threshold is higher for the surface case than for the bulk case. Again, this is a manifestation of the repulsive
nature of the system surface in one-dimension.
\begin{figure}[h]
\noindent\includegraphics[scale=0.55]{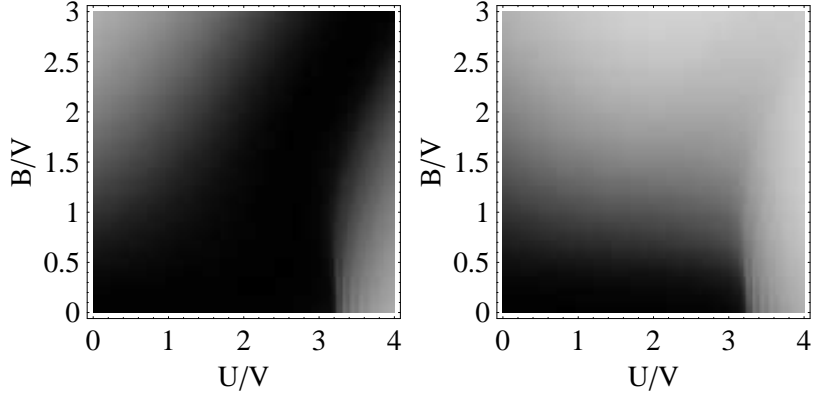}
\caption{Magnetic impurity in ``bulk'': Left (right): Time-averaged selftrapped fraction of
spin up (down) remaining on the initial site, as a function of $U$ and $B$. Dark (white) shade denote selftrapped
fraction close to zero (one).}
\label{fig7}
\end{figure}
\begin{figure}[h]
\noindent\includegraphics[scale=0.55]{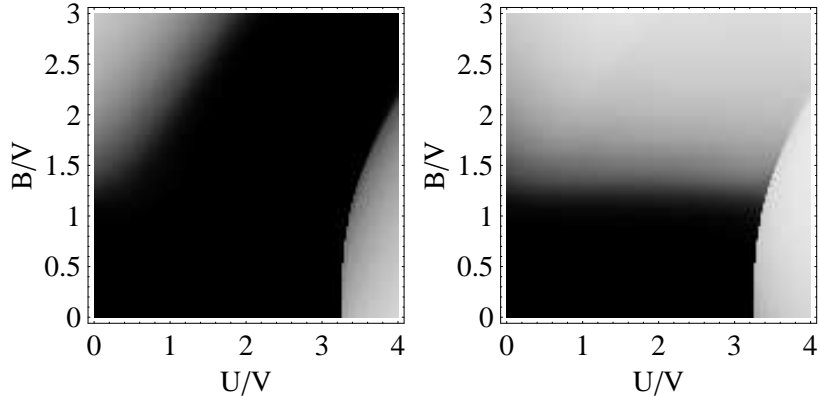}
\caption{Magnetic impurity on surface: Left (right): Time-averaged selftrapped fraction of
spin up (down) remaining on the initial site, as a function of $U$ and $B$. Dark (white) shade denote selftrapped
fraction close to zero (one).}
\label{fig8}
\end{figure}

\section{Conclusions}
We have examined theoretically the formation of bound spin states on top of a narrow magnetic impurity region
in the presence of an external magnetic field, in the simplified framework of modelling the electronic 
repulsion by means of a local nonlinear term. The ensuing set of coupled DNSL equations obtained for the 
vector impurity, is solved numerically for arbitrary values of electronic repulsion and magnetic
field strength. We focused on the cases of a bulk and a surface impurity. In general, we found that it is easier
to create a bound state in the bulk case than in the surface case. In the latter, the bound state diagram showing the
number of bound states in terms of nonlinearity and magnetic field parameters, show a rather complex structure, that
includes a ``forbidden region'' delimited approximately by $U=B+V$ and $U=B-V$. In this regard, the presence of a 
surface inhibits completely the creation of a vector localized mode, in the region $U\sim B$.  In addition, there is a 
parameter region where two bound states exist, although only one of them is (linearly) stable. 
Dynamical results for selftrapping show that, in general, it is easier to selftrap on a bulk impurity than on a 
surface one, although the selftrapping is sharper for the latter. This repulsive character of the surface was observed earlier for  the one-dimensional nonlinear 
chain\cite{mvk_OL}, and seems to be generic to  one-dimensional discrete systems.

\section{Acknowledgements}

This work has been supported by Fondecyt Grant 1080374.

\newpage

\section*{\centerline{List of Figure Captions}}

\noindent
Figure 1:\ \ Left: $B=0$ case. Amplitude of localized state for both spins as a function of nonlinearity.
Right: $U=0$ case. Amplitude of localized state for both spins, as a function of magnetic field. 
\vspace{0.4cm}

\noindent
Figure 2:\ \ Amplitude of localized mode as a function of nonlinearity and
external magnetic field, for spin up (left) and spin down (right). Darkest (whitest) shade 
corresponds to $\xi=-1(1)$. 
\vspace{0.4cm}

\noindent
Figure 3:\ \ (Color online). Spatial mode profiles for $U/V=3$ and varying values
of $B/V$. Top left: $B/V=1$. Top right: $B/V=2$. Bottom left: $B/V=4$. Bottom right: $B/V=6$. 
Solid (dashed) curves denote the spin up (down) mode.
\vspace{0.4cm}

\noindent
Figure 4:\ \  Left: $B=0$ case. Amplitude of localized state for both spins as a function of nonlinearity.
Right: $U=0$ case. Amplitude of localized state for both spins, as a function of magnetic field. 
\vspace{0.4cm}

\noindent
Figure 5:\ \ Number of localized surface modes, as a function of nonlinearity and
magnetic field.
\vspace{0.4cm}

\noindent
Figure 6:\ \  (Color online). Examples of surface localized modes. Solid (dashed) lines refer to spin up (down). 
Top left: $B/V=1, U/V=4$ (stable mode). Top
right: $B/V=1,U/V=4$ (unstable mode). Bottom left: $B/V=3, U/V=5$. Bottom right: $B/V=3, U/V=1$
\vspace{0.4cm}

\noindent
Figure 7:\ \ (Color online). ``Bulk" magnetic impurity: Time-averaged selftrapped fraction for both spins remaining 
on the initial site, as a function of $U$ and $B$.
\vspace{0.4cm}

\noindent
Figure 8:\ \ (Color online). Surface magnetic impurity: Time-averaged selftrapped fraction for both spins 
remaining on the initial site, as a function of $U$ and $B$.

\end{document}